\documentclass[]{aa}
\usepackage{graphicx}
\usepackage{natbib}
\usepackage[varg]{txfonts}

\begin{document}

\title{Orphan penumbrae: Submerging horizontal fields}
\titlerunning{Orphan penumbrae: Submerging horizontal fields}

\author{J. Jur\v{c}\'{a}k
        \inst{1}
        \and
        L.R. Bellot Rubio
        \inst{2}
        \and
        M. Sobotka
        \inst{1}}

\institute{Astronomical Institute of the Academy of Sciences, Fri\v{c}ova
  298, 25165 Ond\v{r}ejov, Czech Republic
  \and
  Instituto de Astrof\'{\i}sica de Andaluc\'{\i}a (CSIC), Apdo.\ Correos 3004, 18080 Granada, Spain}

\date{Draft: February 26, 2014}

\abstract
  {}
  {We investigate the properties of orphan penumbrae, which are photospheric
    filamentary structures observed in active regions near polarity
    inversion lines that resemble the penumbra of regular sunspots but
    are not connected to any umbra.}
  {We use Hinode data from the Solar Optical Telescope to determine
    the properties of orphan penumbrae. Spectropolarimetric data are
    employed to obtain the vector magnetic field and line-of-sight
    velocities in the photosphere. Magnetograms are used to study the overall
    evolution of these structures, and G-band and \ion{Ca}{II}~H
    filtergrams are to investigate their brightness and apparent
    horizontal motions. }
 {Orphan penumbrae form between regions of opposite polarity in places
   with horizontal magnetic fields. Their magnetic configuration is
   that of $\Omega$-shaped flux ropes. In the two cases studied here,
   the opposite-polarity regions approach each other with time and the
   whole structure submerges as the penumbral filaments disappear.
   Orphan penumbrae are very similar to regular
   penumbrae, including the existence of strong gas flows. Therefore,
   they could have a similar origin. The main difference between them
   is the absence of a ``background'' magnetic field in orphan
   penumbrae. This could explain most of the observed
   differences.}
  {The fast flows we detect in orphan penumbrae may be caused by
  the siphon flow mechanism. Based on the similarities between orphan
  and regular penumbrae, we propose that the Evershed flow is also a
  manifestation of siphon flows.}
 

\maketitle

%
%

\section{Introduction}
\label{introduction}

All structures seen in the solar photosphere other than granulation
are caused by magnetic fields that influence the convection and thus
the presence and shape of granules. The largest, strongest, and most
complex magnetic fields are located in active regions. The
fields can completely inhibit convective motions there, and thus create
sunspots and pores \citep[see][]{Keppens:1996, Solanki:2003,
Schlichenmaier:2009, Borrero:2011}. There are other
structures observed in active regions that are associated with
the submergence or emergence of magnetic flux tubes but not so thoroughly
studied as the more prominent sunspots and pores.

In this paper, we analyse filamentary structures observed in active
regions near polarity inversion lines. Since they visually resemble a
penumbra without an umbra, we call them ``orphan penumbrae'' following
\citet{Zirin:1991}. Similar structures were described before.  There
are numerous reports of alignments of granules or intergranular lanes
near polarity inversion lines of active regions \citep[first
by][]{Miller:1960}. \citet{Loughhead:1961} interpreted the alignments
as being due to magnetic loops rising through the solar photosphere.
The same conclusion was reached by other authors, although the
reported lifetimes and sizes of the structures differ considerably
\citep{Brants:1985, Tarbell:1990, Wang:1992, Schlichenmaier:2010}. In
a recent work, \citet{Kuckein:2012} suggested that orphan penumbrae
are the photospheric signatures of low-lying flux ropes.
However, as pointed out by \citet{Lim:2013} and
  \citet{Zuccarello:2013}, other mechanisms can also lead to the
  formation of orphan penumbrae.  These authors observed orphan
  penumbrae in regions with overlaying chromospheric fields and
  interpreted them as a consequence of rising $\Omega$-loops that are
  trapped in the photosphere by the latter.  Here, we
  suggest that the $\Omega$-loops responsible for orphan penumbrae
  submerge again during their decay phase.

The presence of inclined magnetic fields in the photosphere may indeed
explain the formation of orphan penumbrae from at least a theoretical perspective.
\citet{Chandrasekhar:1961} showed that convection in inclined magnetic
fields is modulated by the strength of the vertical field component,
while the horizontal component determines the shape of the convective
cells. This was observationally supported by \citet{Jurcak:2011},
assuming that penumbral filaments are of convective origin. Numerical
simulations also confirm the importance of the horizontal component of
the magnetic field for the convective cell shape
\citep[see][]{Thomas:2008}. If the magnetic field is horizontal,
convection will not be stopped but the cells will be highly elongated
and resemble filaments, which perhaps, explains the existence of
orphan penumbrae.

Here, we determine the physical properties of orphan penumbrae
during their decay phase using spectropolarimetric
measurements, magnetograms, and G-band and \ion{Ca}{II}~H filtergrams
taken by the Hinode satellite.  Section~\ref{observations} describes
the observations and data analysis.  The magnetic and dynamic
configuration of orphan penumbrae is inferred and compared with that
of regular penumbrae in Sect.~\ref{results}.  The evolution,
brightness, and horizontal motions of these structures are also
discussed there. We interpret the various observational results in
Sect.~\ref{discussion} and summarise our conclusions in
Sect.~\ref{conclusions}.

\begin{figure*}[!t]
 \centering
 \includegraphics[width=0.96\linewidth]{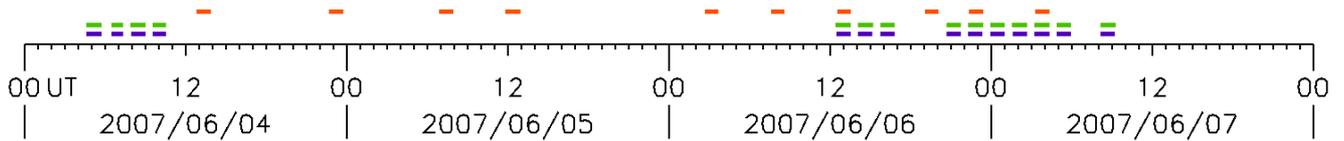}
 \caption{Chart showing the observations of active region
 NOAA~10960. The blue and green dashes indicate BFI observations in
 G-band and \ion{Ca}{II}~H, respectively. The red dashes indicate
 spectropolarimetric scans. The arrows mark the data shown and discussed
in this paper.}  \label{time_axis_2007} 
\end{figure*}

\begin{figure*}[!t]
 \centering
 \includegraphics[width=0.96\linewidth]{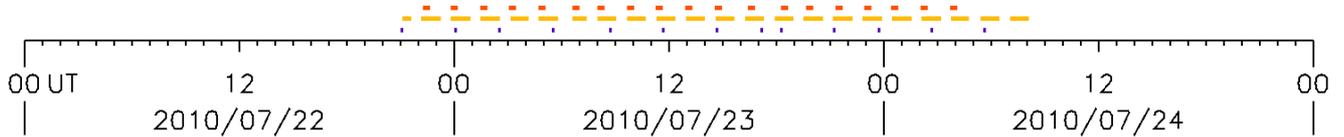}
 \caption{Analogous to Fig.~\ref{time_axis_2007} but for active
 region NOAA~11089. The yellow colour indicates NFI V/I~magnetograms
 in the \ion{Na}{I}~D. All the NFI V/I~magnetograms are
 used in this paper, so they are not marked individually by arrows.}
 \label{time_axis_2010}
\end{figure*}

\section{Observations and data analysis}
\label{observations}

We use data obtained with the three instruments of the Solar Optical
Telescope \citep[SOT,][]{Tsuneta:2008} aboard the Hinode satellite
\citep{Kosugi:2007}, namely, the Broad-band Filter Imager (BFI), the
Narrow-band Filter Imager (NFI), and the Spectropolarimeter (SP). The
BFI data were taken through a G-band filter centred at 430.5~nm with
a bandpass of 0.8~nm and through a \ion{Ca}{II}~H filter centred at
396.85~nm with a bandpass of 0.3~nm. The pixel sampling corresponds to
0\farcs11 (2$\times$2 binning), which results in a diffraction-limited
spatial resolution of 0\farcs22. The NFI measurements were acquired in
the red wing of the \ion{Na}{I}~D line with the polarisation unit to
obtain V/I magnetograms at a spatial resolution of 0\farcs32 (pixel
sampling of 0\farcs16). The SP observed the Stokes profiles of the two
\ion{Fe}{I} lines at 630.15 and 630.25~nm.  The SP measurements were
taken in the so-called fast mode, which is when the pixel sampling is 0\farcs32,
the exposure time is 1.6~s, and the resulting noise level is
$10^{-3}I_{\rm c}$. The data were calibrated with the standard routines
available in the Hinode SolarSoft package.  A subset of these
observations has been analysed by \citet{Zuccarello:2013}.

\begin{figure*}[!t]
 \centering \includegraphics[width=0.95\linewidth]{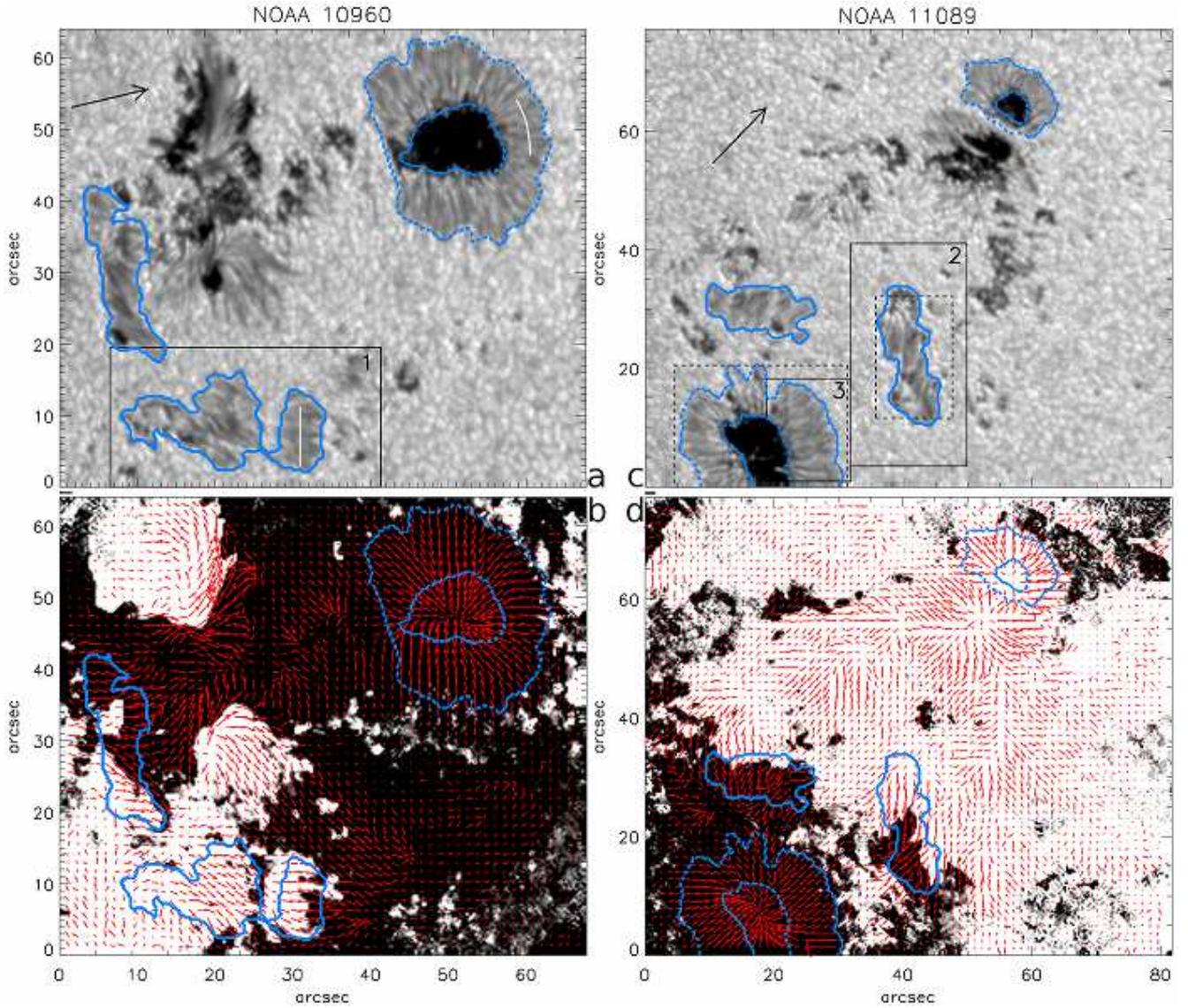}
 \caption{Continuum intensity maps showing an active region NOAA~10960 on
 4 June 2007 around 13~UT (a) and NOAA~11089 on
 22 July 2010 around 22:30 UT (c) at heliocentric
 angles of 47$^\circ$ and 43$^\circ$, respectively.  North is up and
 east to the left.  Maps of the vertical component of the magnetic
 field in the LRF saturated at $\pm 25$~G are shown in (b)
 and (d). The red arrows mark the strength and orientation of the
 horizontal component of the magnetic field. The length of the thick
 lines above the maps (b) and (d) correspond to 1000~G. The solid and
 dashed contours encircle the orphan and regular penumbrae,
 respectively. The boxes indicated by solid black lines mark the areas
 displayed in Figs.~\ref{ambiguity} and~\ref{detail_op}. The
 boxes indicated by dashed black lines mark the areas displayed in
 Fig.~\ref{aa_maps}. The white lines in (a) mark the positions of cuts
 displayed in Fig.~\ref{cuts}.}  \label{magnetogram}
\end{figure*}

According to \citet{Shelyag:2004}, the G-band forms at around
$\tau_{500}=-1$, which is comparable to the formation height of 100~km
reported by \citet{Jess:2012}. Since the typical granulation pattern
is observed in the G-band images, the formation height has to be below
140~km, which is the boundary between normal and reverse granulation
\citep{Cheung:2007}. Because of the width of the Hinode \ion{Ca}{II}~H
filter and the broad response/contribution functions of the line
(as computed for the FALC model by \citealt{Carlsson:2007} or
\citealt{Rezaei:2008}), the formation height of the \ion{Ca}{II}~H images
is not well defined. They mostly sample photospheric layers
below 300~km (reverse granulation pattern in quiet Sun regions), but
there are also contributions from the chromosphere. The
Hinode \ion{Ca}{II}~H filtergrams show strong emission at the position
of enhanced magnetic activity. The \ion{Na}{I}~D magnetograms were
taken $+14$~pm from the line core.  This part of the line wing is
formed around 200~km above the solar surface \citep{Leenaarts:2010}.
According to \citet{Cabrera:2005}, the two iron lines recorded by the
SP are most sensitive to layers around $\tau_{500}=-1$. The physical
parameters obtained from the inversion of the \ion{Fe}{I} lines,
therefore, give us information about this atmospheric height.

The observations used here correspond to two active regions
(NOAA~10960 and 11089) and are summarised in
Figs.~\ref{time_axis_2007} and~\ref{time_axis_2010}. In both cases, we
did not capture the formation of the orphan penumbrae. We observed
their evolution and decaying phase only. The data marked with
arrows in Figs.~\ref{time_axis_2007} and~\ref{time_axis_2010} are
presented in detail in this paper. However, all the measurements were
analysed to study the evolution of the orphan penumbrae and to confirm
the validity of our results.

In the case of NOAA~10960 (Fig.~\ref{time_axis_2007}), there are
unevenly separated SP scans of the active region, and the orphan
penumbra is located on the southern edge of the field of view
(FOV). There is a two-day gap in the BFI data as the structure was out
of the FOV of this instrument. In the initial sequence of BFI images
taken on 4 June 2007, the cadence of the filtergrams is 60~s
for both datasets, while it is 100~s for
the G-band images and 20~s for the \ion{Ca}{II}~H filtergrams in the rest of the BFI data. The active
region was located between [47$^\circ$E, 6$^\circ$S] and [6$^\circ$E,
6$^\circ$S] during the analysed time period (heliocentric angles
between 48$^\circ$ and 6$^\circ$).

We have at our disposal SP scans of active region NOAA~11089 that repeat
approximately every 100~minutes (Fig.~\ref{time_axis_2010}).
Moreover, there are almost uninterrupted observations of Na~I~D~V/I
magnetograms with a cadence of 5 minutes. However, the photospheric
evolution of NOAA~11089 is not well covered, since only 13 BFI images
were taken through the G-band filter. This active region moved from
[31$^\circ$E, 27$^\circ$S] to [14$^\circ$E, 28$^\circ$S] during the
observations (heliocentric angles between 43$^\circ$ and 32$^\circ$).

\begin{figure*}[!t]
 \centering 
 \includegraphics[width=0.69\linewidth]{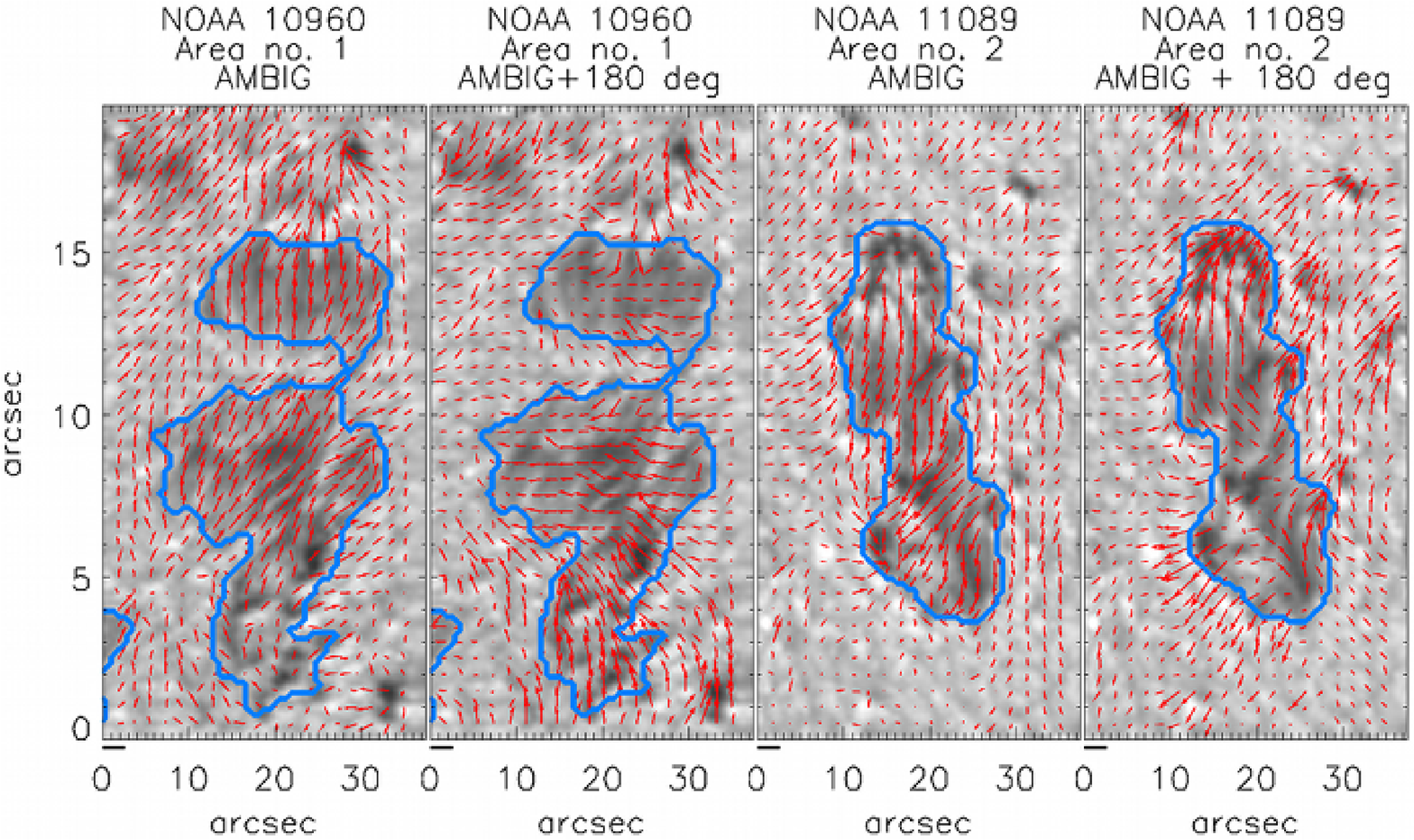}
 \caption{Continuum intensity maps showing the orphan penumbrae in
 areas no.~1 and~2 of Fig.~\ref{magnetogram}. The arrows represent the
 orientation and strength of the horizontal component of the vector
 magnetic field in the LRF. For each of the areas, we display the two
 LRF azimuths that result from the two possible LOS azimuth
 solutions. The left panels show the correct solution. The lengths of
 the thick lines near the beginning of the $x$~axes correspond to
 1000~G.  Area no.~1 is rotated by 90$^\circ$ for display purposes.}
 \label{ambiguity}
\end{figure*}

\begin{figure*}[!p]
 \centering 
 \includegraphics[width=0.91\linewidth]{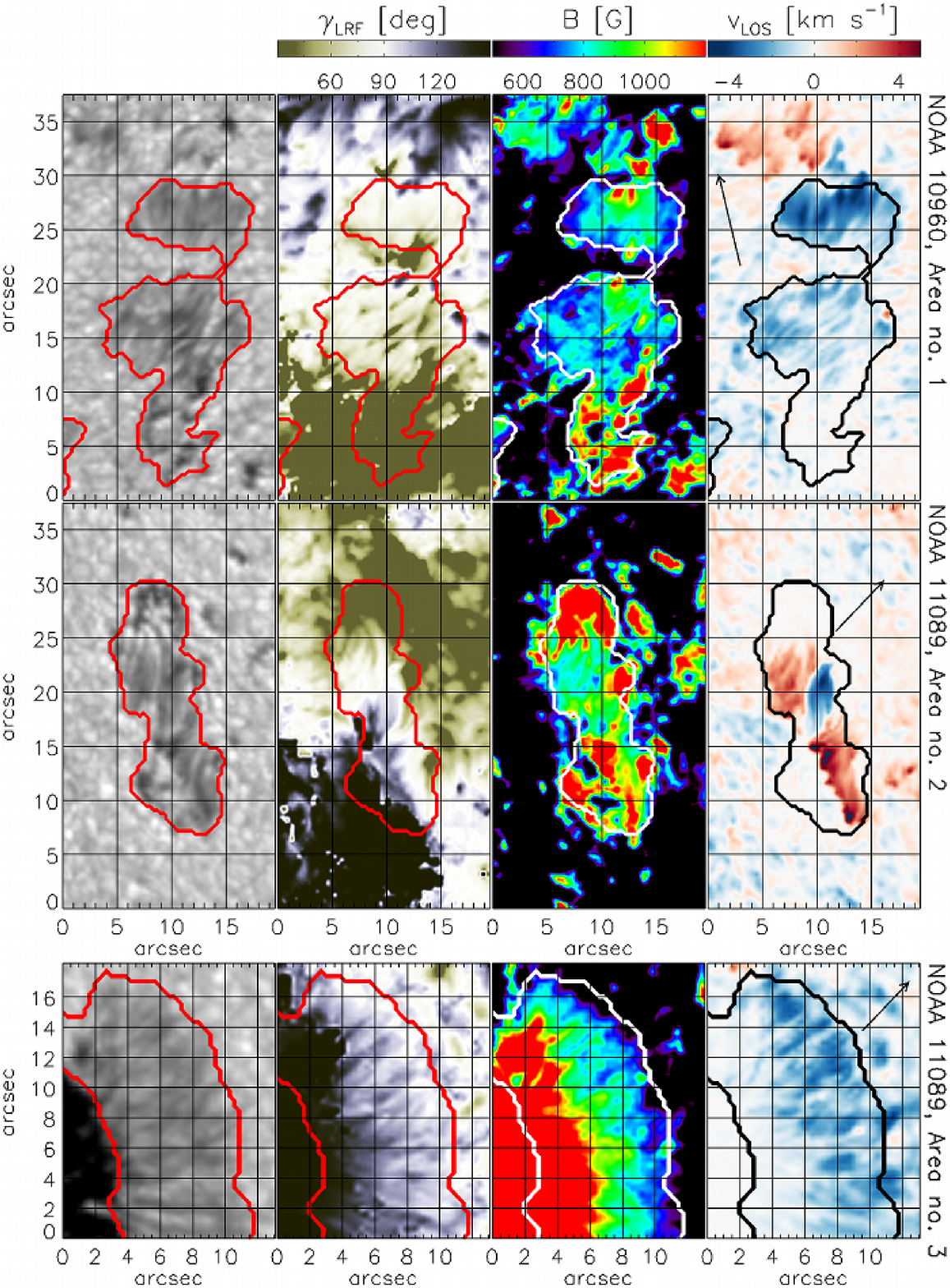}
 \caption{Left
 to right: Maps of continuum intensity, LRF magnetic field inclination
 ($\gamma_\mathrm{LRF}$), magnetic field strength ($B$), and LOS~velocity
 ($v_\mathrm{LOS}$) in the areas marked by solid black boxes in
 Fig.~\ref{magnetogram}. Area no.~1 is rotated by 90$^\circ$ for
 display purposes. The contours are the same as in
 Fig.~\ref{magnetogram}.  Areas no.~1 and no.~2 show the properties of
 orphan penumbrae segments and area no.~3 of the centre side of
 regular penumbra. The arrows in the LOS~velocity maps point towards
 the disc centre.}  \label{detail_op}
\end{figure*}

\subsection{Data analysis}
\label{analysis}

We used the SIR inversion code \citep[Stokes Inversion based on
Response functions;][]{Cobo:1992} to retrieve the values of the
atmospheric parameters from the Stokes profiles observed by the Hinode
SP. The pixel sampling of 0\farcs32 is comparable to the width of the
studied filaments, so they are not spatially resolved. For this
exploratory analysis, we adopted a simple model atmosphere assuming
that the physical conditions do not change with optical depth with the
exception of temperature (We allowed for three optical depths at which
the temperature can change and obtained the stratification in a finer 
grid by interpolation.).  We took into account the spectral point
spread function of the Hinode SP, allowed for micro-turbulent
velocity, and set the macro-turbulent velocity to zero. We did not use
any stray light profile.

The magnetic field inclinations and azimuths retrieved from the
inversion are expressed in the line-of-sight (LOS) reference frame.
We applied the code AMBIG \citep{Leka:2009} to solve the 180$^\circ$
ambiguity of the LOS azimuth. The AMBIG code is based on the minimum energy
method \citep{Metcalf:1994}. The disambiguated LOS vector magnetic
field was then transformed to the local reference frame (LRF) with the
help of routines from the AZAM code \citep{Lites:1995}.

We used local correlation tracking \citep[LCT,][]{November:1988} to
study horizontal motions in the G-band and \ion{Ca}{II}~H
filtergrams. Applied to a time series of images, the LCT technique provides
the time-averaged horizontal velocity field of all structures in the
FOV. We first aligned the BFI images and removed the p-mode
oscillations by applying a $k-\omega$~filter with a cut-off of
5~km~s$^{-1}$. Then, we selected a Gaussian tracking window of FWHM
1\arcsec\/ (9 pixels), which is fine enough to obtain proper motions of
structures within filamentary regions, and the temporal integration
was made over an interval of 18~minutes for both G-band and
\ion{Ca}{II}~H data. The resulting velocity maps do not represent real
gas flows; for example, there are motions of
structures towards the umbra in most penumbrae, although the gas flow is in the opposite
direction \citep{Sobotka:1999}.

\section{Results}
\label{results}

\subsection{Magnetic and velocity structure}
\label{sp_data}

In Fig.~\ref{magnetogram}, we present continuum intensity maps of
active regions NOAA~10960 and 11089 constructed from the SP scans
with solid and dashed contours encircling orphan and regular
penumbrae. To show the location of the orphan penumbrae with respect
to the polarity inversion line (PIL), we created highly saturated maps
of the vertical component of the magnetic field ($\pm 25$~G;
Figs.~\ref{magnetogram}b and~~\ref{magnetogram}d) using the magnetic
field strengths and LRF inclinations derived from the inversion. 

In all cases, the orphan penumbral segments are formed between regions
of opposite polarity and, indeed, most of them lie on PILs. When orphan
penumbrae sit in unipolar areas, the opposite polarities are found in
pore-like structures located at the end of some of the filaments. In
Figs.~\ref{magnetogram}b and~\ref{magnetogram}d, the orientation and
strength of the horizontal component of the vector magnetic field in
the LFR is marked by red arrows. As can be seen, the orphan penumbral
filaments and the magnetic field are aligned and cross the PILs from
the positive to the negative side (normal configuration).  The
filaments of the orphan penumbrae observed by \citet{Kuckein:2012}
were found to be aligned with the magnetic field too. However, they
ran parallel to the PIL instead of crossing it, which was ascribed to
the existence of a chromospheric filament above them.

In Fig.~\ref{ambiguity}, we show the two possible orientations
  of the LRF azimuth for the orphan penumbrae of areas no.~1 and~2 in
  Fig.~\ref{magnetogram}. The solution displayed on the left panels
  corresponds to that determined by the AMBIG code. We believe this is
  the correct solution, because it leads to a magnetic field well
  aligned with the penumbral filaments. Such a solution implies an
  $\Omega$-shaped configuration of the magnetic field lines. The
  solution on the right panels corresponds to the opposite orientation
  of the LOS azimuth and produces a LRF magnetic field that is not
  aligned with the filaments. This scenario does not seem realistic,
  and therefore, we discard it.  All the orphan penumbrae observed
in active regions NOAA~10960 and 11089 are compatible with
$\Omega$-shaped flux ropes.

Figure~\ref{detail_op} displays maps of continuum intensity, magnetic
field inclination in the LRF ($\gamma_\mathrm{LRF}$), magnetic field
strength ($B$), and LOS velocity ($v_\mathrm{LOS}$) as derived from
the inversion of the two areas.  For comparison, we also display a
segment of a regular penumbra using the same colour coding.

From the LRF inclination maps of Fig.~\ref{detail_op}, it is clear
that the orphan penumbral filaments form in
regions where the field is mostly horizontal and that they stretch
across the PIL, confirming our interpretation of the saturated $B_z$
maps of Fig.~\ref{magnetogram}. Their magnetic field inclination is
within $\pm 10^\circ$ from being completely horizontal. Near the
footpoints of the filaments, the field is inclined by $30^\circ$ with
respect to the local normal line. In the centre-side part of the regular
penumbra (area no.~3), the LRF inclination map shows that the field turns
horizontal radially outwards while keeping the polarity of the spot
\citep[as described by numerous authors, see the review
by][]{Solanki:2003}.

The magnetic field strength is around 900~G in orphan penumbrae with
generally weaker fields in area no.~1 compared to area no.~2. Near the
footpoints of the filaments, the field tends to be stronger. A detailed
analysis of the variation of the magnetic field strength along orphan
penumbral filaments is presented in Sect.~\ref{discussion}. In
the regular penumbra, we observe a steady decrease of the field strength
from the umbra towards the outer sunspot edge \citep[][]{Solanki:2003}.

The LOS velocity maps displayed in Fig.~\ref{detail_op} make it clear
that orphan and regular penumbrae have a similar flow structure. The
flows are aligned with the filaments and become faster further from
the location of their origin, as it also happens in sunspots
\citep[see][]{Borrero:2005}. The velocity peaks are
always reached at the filament ends, which allows us to determine the
tails of the filaments. In the case of redshifted flows (away from
the observer), this behaviour can to some extend be caused by
projection effects as the flow turns towards the solar surface and has
a more favourable orientation to the LOS.  However, the
flow gets faster further from the location of their origin also in the case
of flows oriented towards the observer (blueshifted segments), where the
angle between the flow direction and the LOS is larger (less
favourable orientation to the LOS).

A unique property of orphan penumbrae is the existence of flows of
opposite direction in neighboring filaments. Regular penumbrae do not
normally show such oppositely-directed motions, although there are cases of penumbral segments with inward gas
flows during the initial phases of their evolution in the Hinode
data archive. As described
later, the oppositely oriented flows observed in orphan penumbrae
eventually disappear.

\begin{figure*}[!t]
 \centering 
 \includegraphics[width=0.95\linewidth]{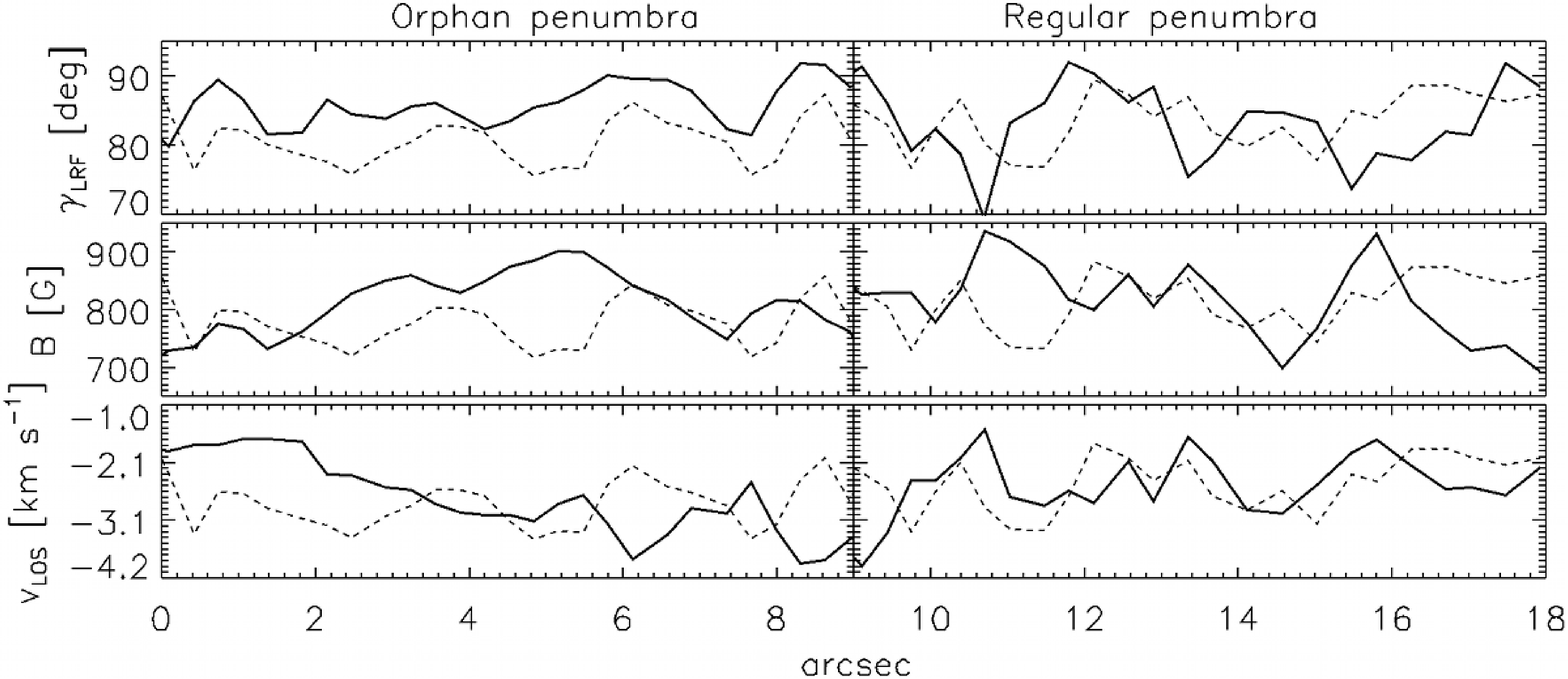}
 \caption{Variations of $\gamma_\mathrm{LRF}$, $B$, and
 $v_\mathrm{LOS}$ along the cuts marked with white lines in
 Fig.~\ref{detail_op}. The solid and dashed lines show the atmospheric
 parameters and the continuum intensity, respectively. For display
 purposes, we changed the polarity of the field along the cut through
 the filaments of the regular penumbra and also increased
 $\gamma_\mathrm{LRF}$ by 10$^\circ$ and $v_\mathrm{LOS}$ by
 1~km~s$^{-1}$ there. The range of continuum intensities used to draw
 the dashed lines is the same for orphan and regular penumbrae.}
 \label{cuts}
\end{figure*}

In Fig.~\ref{cuts}, we compare the variation of parameters 
along cuts perpendicular to filaments in orphan and regular penumbrae
(see white lines in Fig.~\ref{magnetogram}a). Since orphan and regular
penumbrae were observed at the same heliocentric angle and analysed
with the same inversion technique, the magnetic and dynamic parameters
we obtain should be fully comparable.

\begin{figure*}[!t]
 \centering \includegraphics[width=0.95\linewidth]{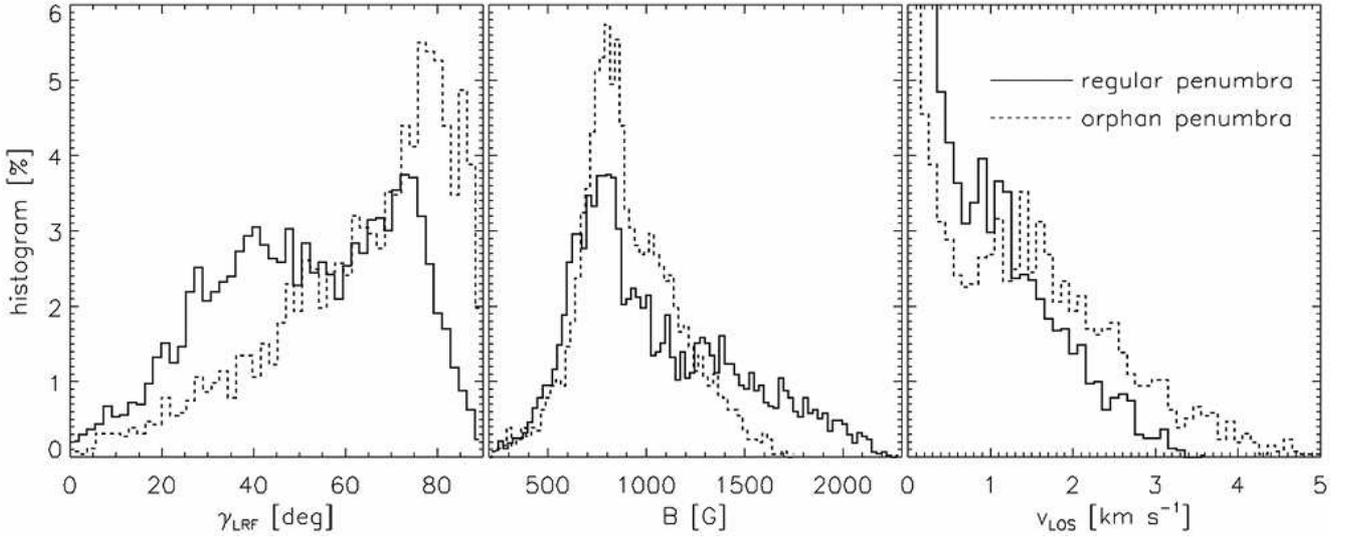}
 \caption{Histograms of the LRF magnetic field inclination
(left;
   90$^\circ$ is horizontal), magnetic field strength (middle), and
   absolute values of the LOS~velocity (right) in regular and
orphan penumbrae.
   The pixels used to create the regular penumbra histograms correspond to those
in the two regular penumbrae marked by dashed lines in Fig.~\ref{magnetogram}c.
The orphan penumbra histograms are based on the pixels in two orphan
penumbrae encircled by solid contours in Fig.~\ref{magnetogram}c.}
 \label{scatter_plot}
\end{figure*}

In the orphan penumbra, there are small variations of 
the LRF inclination across the filaments with maximum changes of
10$^\circ$. In the regular penumbra, bright and dark filaments show
larger differences of up to 20$^\circ$.  The field strength
differences between bright and dark filaments are also larger in the
regular penumbra (more than 100~G) compared to the orphan penumbra,
where there are global changes of $B$, but no pronounced fluctuations
across individual filaments. Likewise, variations of the LOS velocity
are more prominent across regular penumbral filaments (2~km~s$^{-1}$)
than across orphan penumbral filaments (up to 1~km~s$^{-1}$).

We do not observe significant correlations between
continuum intensity and magnetic or dynamic parameters along the cuts
in either orphan or regular penumbrae. However, there is a clear
tendency of faster flows in weaker and more horizontal magnetic fields
in the regular penumbra, which can be interpreted as being due to the
presence of a steady, stronger, and more vertical background magnetic
field. Along the orphan penumbral cut, the faster flows tend to be
located in regions with stronger and more horizontal fields. This is
not compatible with the existence of a background magnetic field. A
detailed analysis of the relation between flow velocity and field
strength in orphan penumbrae is carried out in Sect.~\ref{discussion}.

To compare the global properties of orphan and regular penumbrae, we
present histograms of $\gamma_\mathrm{LRF}$, $B$, and $v_\mathrm{LOS}$
in Fig.~\ref{scatter_plot}. This analysis is based on the segments of
orphan and regular penumbrae marked with solid and dashed contours in
Fig.~\ref{magnetogram}c.

The left panel of Fig.~\ref{scatter_plot} shows the field 
inclination distributions in orphan and regular penumbrae. As
mentioned before, the magnetic field is mostly horizontal in orphan
penumbrae, whereas regular penumbrae show inclinations ranging from
25$^\circ$ to 80$^\circ$. This result can be understood by keeping in
mind that the information about the magnetic field inclination (and
strength) in regular penumbrae is dominated by the sunspot background
field, and our simplified inversion scheme cannot account for the field
component harboring the gas flow \citep[whose inclination increases
with radial distance and even submerges in the mid-penumbra and
beyond, see the review by][]{Borrero:2011}. Pixels with more vertical
magnetic fields are located outside of the filamentary regions of
orphan penumbrae.

In the middle panel of Fig.~\ref{scatter_plot}, we present 
histograms of the field strengths in orphan and regular penumbrae.
Similarly to the field inclination, there is a significantly larger
spread of $B$ in regular penumbrae compared to orphan penumbrae. This
is again likely caused by the behaviour of the background magnetic
field in regular penumbrae, whose strength steadily decreases from the
umbra to the outer sunspot edge. Pixels with $B$ over 1200~G are
located outside of the filamentary parts of orphan penumbrae.

The right panel of
Fig.~\ref{scatter_plot} shows the LOS velocity distributions of both
types of penumbrae. Unlike the magnetic field strength and
inclination, the shapes of the distributions are comparable.  This
could be due to the fact that the background magnetic field component
does not contain fast flows
\citep{Borrero:2004, Bellot:2004} and thus does not 
significantly influence the retrieved values of the LOS velocity, which is
almost exclusively determined by the flow component. Pixels with
LOS~velocities smaller than 0.5~km~s$^{-1}$ are located near the
umbra/penumbra boundary in regular penumbrae and in pore-like
structures near the ends of orphan penumbrae filaments.

The only difference in the flow field of orphan and regular
penumbrae is the amplitude of the LOS~velocity. We found maximum
LOS~velocities of 5~km~s$^{-1}$ and 3.7~km~s$^{-1}$, respectively. 
This is to some extent due to projection effects,
because we observe only blueshifted regions on the centre side of
regular penumbrae (like in area no.~3). As mentioned before, the orientation of the
flow with respect to the LOS is more favourable in redshifted regions
than in blueshifted regions near the
filaments tails, where the flow is fastest.  However, the flows observed in the
blueshifted regions of the orphan penumbrae are also faster than those
detected in the regular penumbra. This may point to a different
submergence angle of the filaments in the two types of structures.

\subsection{Stokes profile asymmetries}

\begin{figure}[!t]
 \includegraphics[width=1\linewidth]{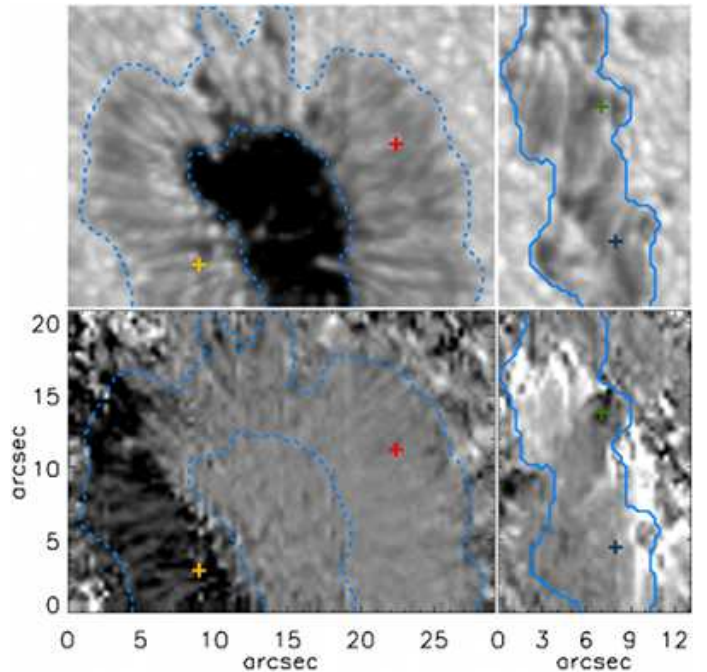}
 \caption{Continuum intensities (top) and Stokes V area asymmetries
 (bottom) within the dashed black boxes of Fig.~\ref{magnetogram}c. The
 area asymmetry maps are saturated at $\pm 0.5$. The coloured $+$~symbols
 mark the positions of the Stokes profiles shown in Fig.~\ref{aa_prof}.}
 \label{aa_maps}
\end{figure}

Although they are not common, highly asymmetric Stokes profiles
can be observed in orphan penumbrae. This is demonstrated by the maps
of Stokes V area asymmetry $\left( \sum_\lambda V/\sum_\lambda
|V|\right)$ of Figure~\ref{aa_maps}. 

\begin{figure*}[!t]
 \includegraphics[width=1\linewidth]{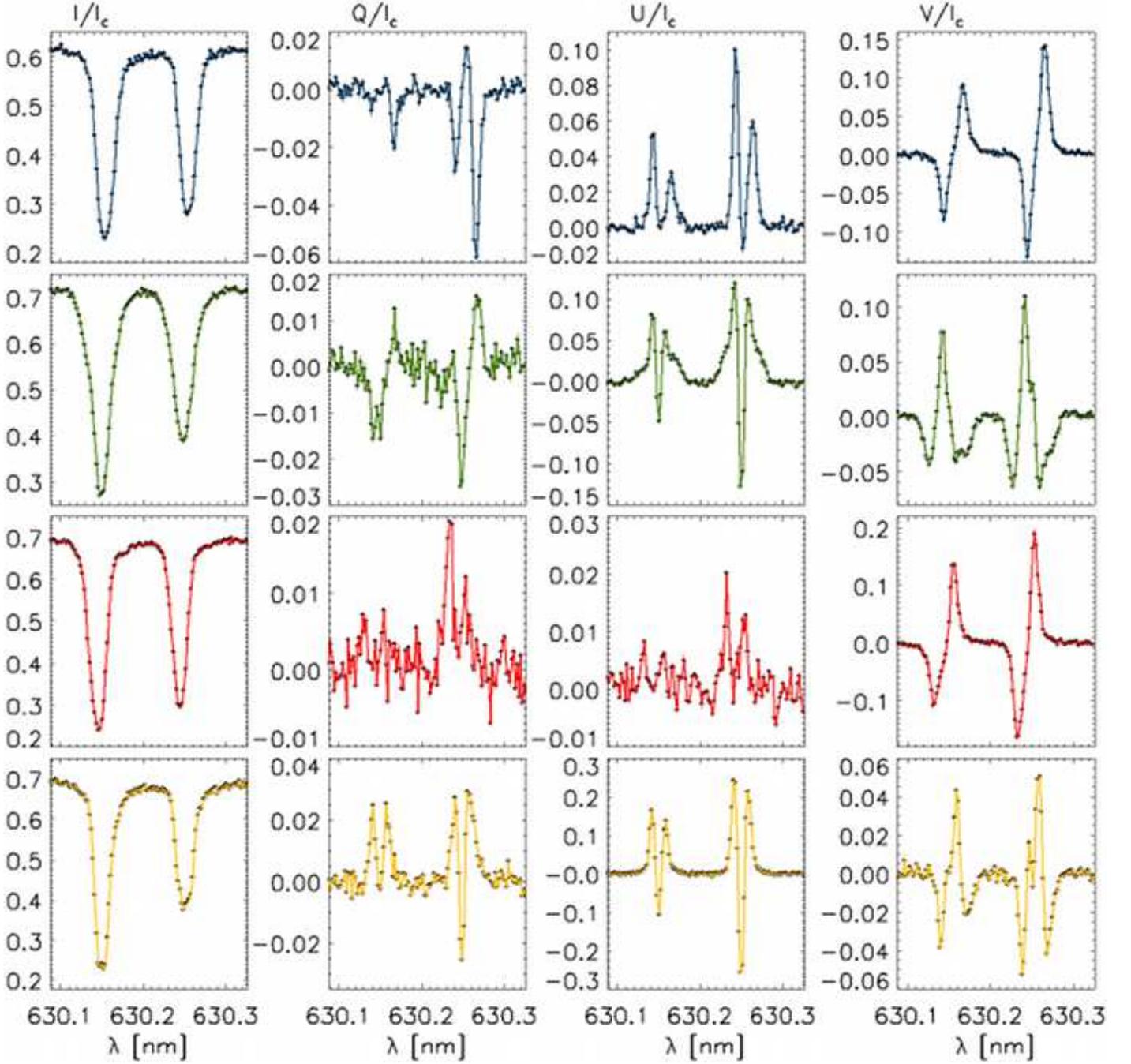}
 \caption{Stokes profiles observed in the pixels marked with coloured
 $+$~symbols in Fig.~\ref{aa_maps}. From top to bottom, we show
 symmetric and asymmetric profiles in the orphan penumbra, along with 
 symmetric and asymmetric profiles in the regular penumbra.} \label{aa_prof}
\end{figure*}

Very asymmetric Stokes V~profiles tend to occur at the border of the
filamentary region of the orphan penumbra, while there are only small
asymmetries elsewhere. The centre side of the regular penumbra also
shows mild asymmetries, whereas the limb side exhibits stronger
asymmetries, although the magnetic configuration is the
same on both penumbral sides. In Fig.~\ref{aa_prof}, we present
examples of observed Stokes profiles. In orphan and regular penumbrae,
one can find highly asymmetric Stokes V profiles (second and fourth
rows, respectively) but also nearly symmetric profiles (first and
third rows).

\citet{Zuccarello:2013} interpreted the Stokes V asymmetries of orphan
penumbrae as a proof of the existence of a background magnetic field.
However, this is not the only mechanism that can produce asymmetric
Stokes profiles. Strong asymmetries are observed in
canopy-like regions, where strong vertical gradients of the physical
parameters occur \citep{Ishikawa:2010, Viticchie:2012,
  SainzDalda:2012}. The edges of orphan penumbrae show the largest
asymmetries and are the places, where the configurations described by
\citet{Ishikawa:2010} and \citet{Viticchie:2012} may be more relevant
(albeit of much larger size), given that the horizontal field
has to bend rapidly to become more vertical and stronger at the
footpoints of the $\Omega$-shaped loop that forms the penumbra. These
are the places where the line of sight encounters the strongest
vertical gradients, which would produce asymmetric Stokes profiles
without the intervention of any background magnetic field. To estimate more precisely the configuration of magnetic
field in these regions, we would have to employ a depth-dependent inversion of the observed Stokes profiles. This is beyond
the scope of this paper.

Thus, we conclude that the absence of large Stokes asymmetries in the
central part of orphan penumbrae is compatible with the absence of a
background magnetic field in those structures.

\begin{figure}
 \centering \includegraphics[width=1\linewidth]{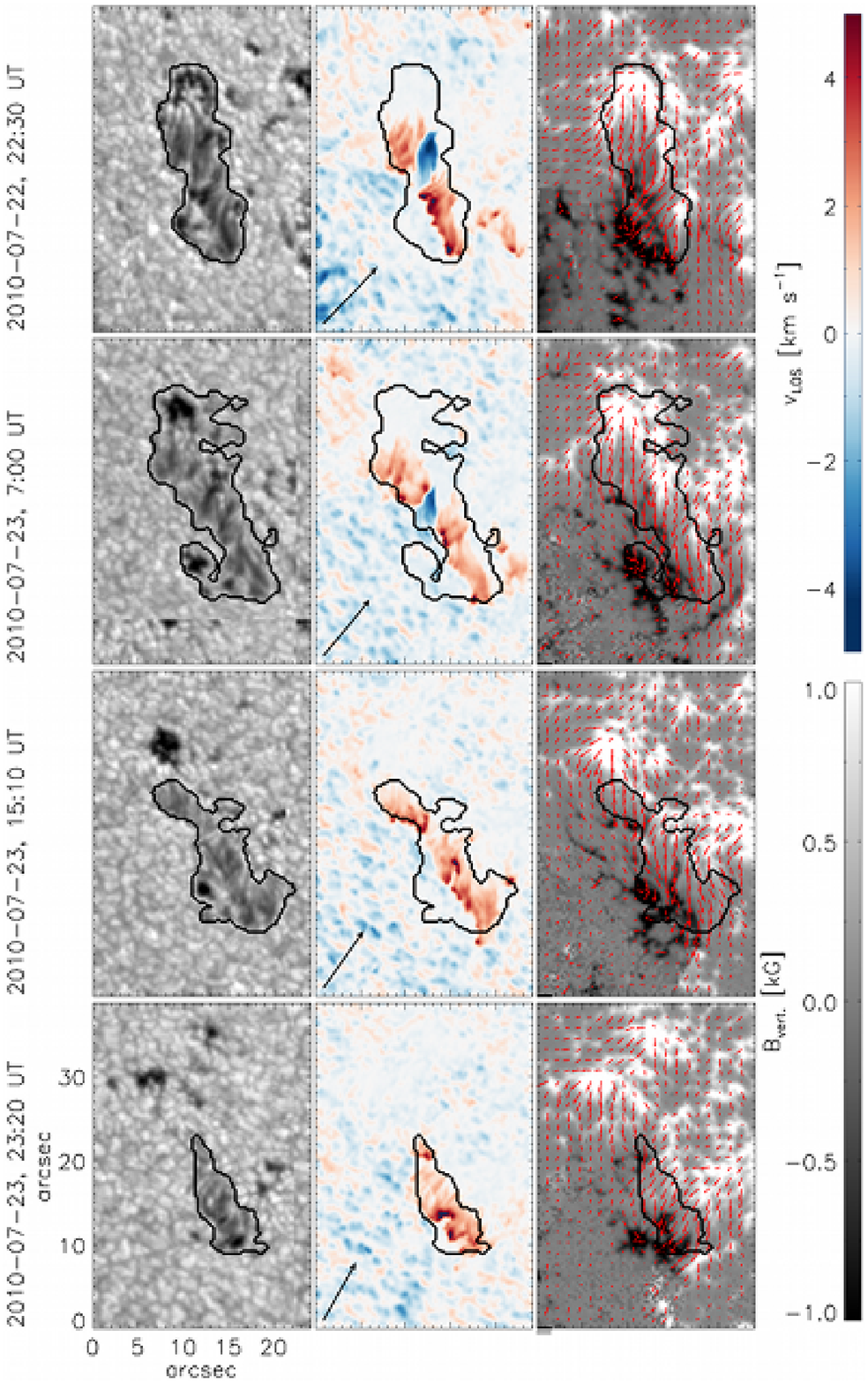}
 \caption{Temporal evolution of the orphan penumbra observed in
 NOAA~11089 (area no.~2 in Fig.~\ref{magnetogram}c).  From left to
 right, we show continuum intensity maps, LOS~velocity maps, and maps
 of the vertical component of the magnetic field with red arrows
 indicating the strength and orientation of the horizontal component.
 The black arrows in the LOS~velocity maps point towards the
 disc centre.}  \label{sp_evol}
\end{figure}

\subsection{Time evolution}
\label{na_mag}

Figure~\ref{sp_evol} shows the temporal evolution of the orphan
penumbra of NOAA~11089 (area no.~2) over 25 hours. In the continuum
intensity maps, the penumbral filaments are observed to get shorter
and the whole structure to decrease in size with time. In the
LOS~velocity maps, the patch with oppositely oriented flow disappears.
As the orphan penumbra approaches the central meridian, the
LOS~velocities decrease. This is not immediately apparent because
the maximum velocities exceed 5~km~s$^{-1}$ in all panels, but there
is a decrease of 0.4~km~s$^{-1}$ of the median LOS~velocity in the
filaments between the first and last maps.

In the magnetic field panels of Fig.~\ref{sp_evol}, one can see a
shear movement of the opposite polarities. The negative polarity
patch in the lower part moves westward (to the right) compared to the
main positive polarity patch. Therefore, the filaments are oriented in
the north/south direction in the first frame, while 
they are oriented in the north-east/south-west direction in the last frame. The upper
segment of the orphan penumbra disappears, as the negative polarity
patch originally located around [10, 20]~arcsec disappears.
This also happens later for the lower segment of the orphan penumbra
when the negative polarity region disappears (see Sect.~\ref{na_mag}
below).

Additionally, we have tracked the evolution of the LOS~magnetic field in
NOAA~11089 for approximately 35 hours using Hinode Na~I~D~V/I
magnetograms. From the centres of gravity of the sunspot
  umbrae, we find that the opposite-polarity leader and follower
spots of NOAA~11089 separated with an average velocity of
130~m~s$^{-1}$, which is larger than the mean values found by
\citet[][80~m~s$^{-1}$]{Sobotka:2007} and
\citet[][45~m~s$^{-1}$]{Svanda:2009}. The leading negative polarity
region was always highly structured, while the number density of
magnetic features around the PIL decreased with time.

\begin{figure*}[!t]
 \centering \includegraphics[width=1\linewidth]{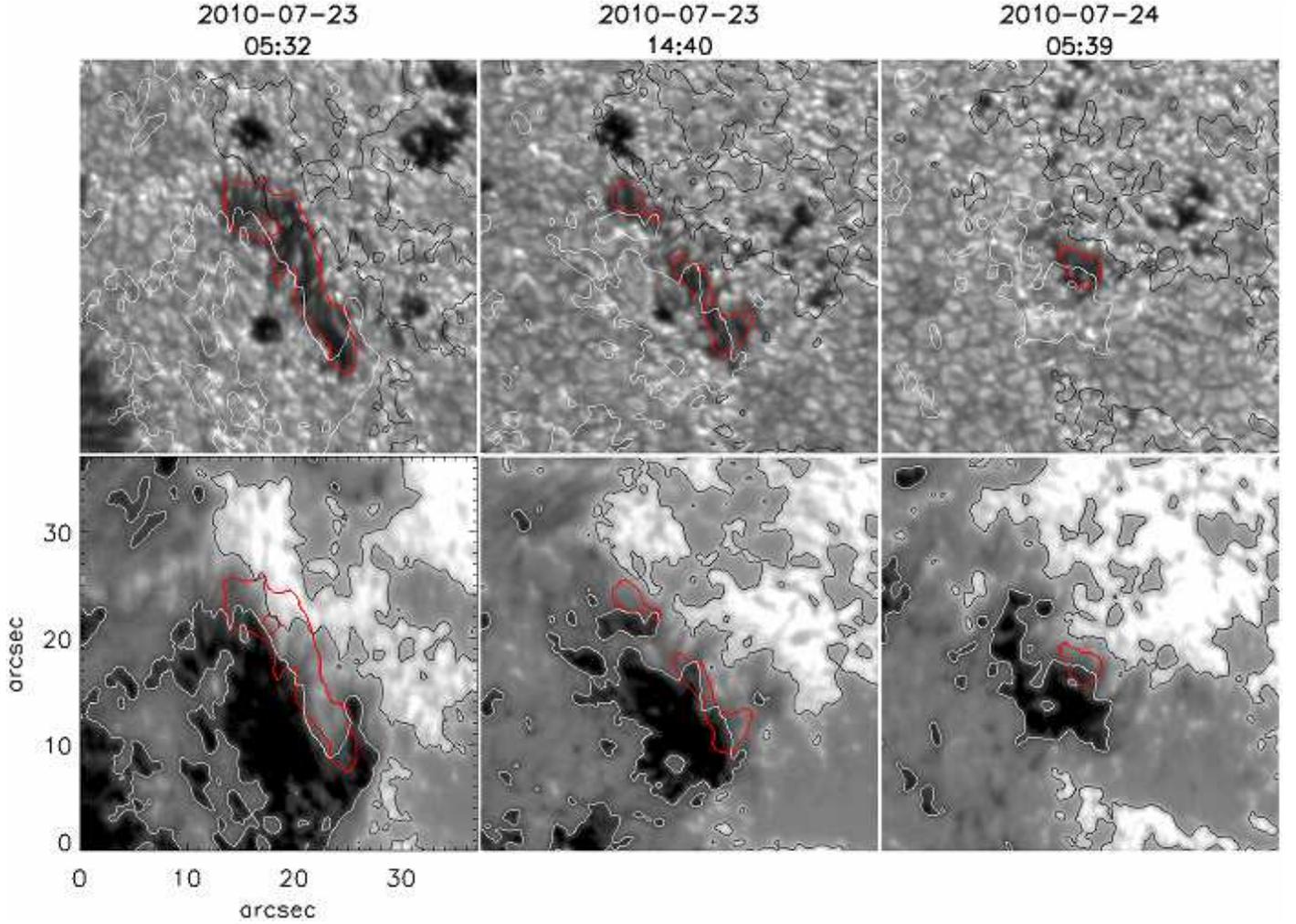}
 \caption{Co-spatial and co-temporal G-band images and \ion{Na}{I}~D
 V/I magnetograms showing the temporal evolution of the orphan penumbra
 observed in active region NOAA~11089 (area no.~2 in
 Fig.~\ref{magnetogram}c). The red contours mark the orphan penumbra location. The black and white contours show the patches
of positive and negative polarity, respectively. The temporal evolution of the V/I magnetograms (lower panels) is
available in the online edition.} \label{nfi_mag}
\end{figure*}

Figure~\ref{nfi_mag} displays three co-spatial and co-temporal G-band
images and \ion{Na}{I}~D~V/I magnetograms summarizing the evolution of
the orphan penumbra. It is clear that the penumbral filaments form
between regions of opposite polarity, confirming the results presented
in Section~\ref{sp_data}. The filaments are located in regions
  with weaker magnetogram signals than the surroundings. The
differences between the \ion{Na}{I}~D~V/I magnetograms of
Fig.~\ref{nfi_mag} and the magnetic field maps of Fig.~\ref{sp_evol}
are caused mainly by the switch from the LOS frame to the LRF.

Played as a movie (see the electronic version of the journal), the
\ion{Na}{I}~D~V/I magnetograms show the opposite-polarity regions
approaching each other until the negative one disappears.  There
  is a significant change in the size of the negative polarity patch,
  which is also clearly seen in Fig.~\ref{nfi_mag}. The positive
  polarity patches next to the orphan penumbra change their shapes
  with time but are preserved due to nearby pores that trail the
  leading sunspot. As there are no stable structures in the two
  regions, it is difficult to calculate their approach velocity. From
  the mean distance between the opposite-polarity patches (determined
  manually), we estimate the approach velocity to be around
  40~m~s$^{-1}$. We obtained a similar velocity of 30~m~s$^{-1}$ by
  tracking the evolution of the length of the orphan penumbral
  filaments in the sequence of G-band images.

Since the two polarities approach with time and one polarity
  disappears according to the \ion{Na}{I}~D~V/I magnetograms, the
  $\Omega$-shaped magnetic field responsible for the orphan penumbra
  must be submerging. The fragmented structure of the patches and
  their evolution also suggest that we do not observe a monolithic
  flux tube but a number of smaller tubes behaving in a similar way.
  There is also the possibility that the magnetic field in the
  negative polarity region dissipates, while it approaches the positive
  polarity region. However, we consider this a less plausible scenario
  because dissipating polarities in active regions tend to show increasing
  distances instead \citep{Driel:1990, Moreno:1994}.

\begin{figure*}[!t]
 \centering 
 \includegraphics[width=0.99\linewidth]{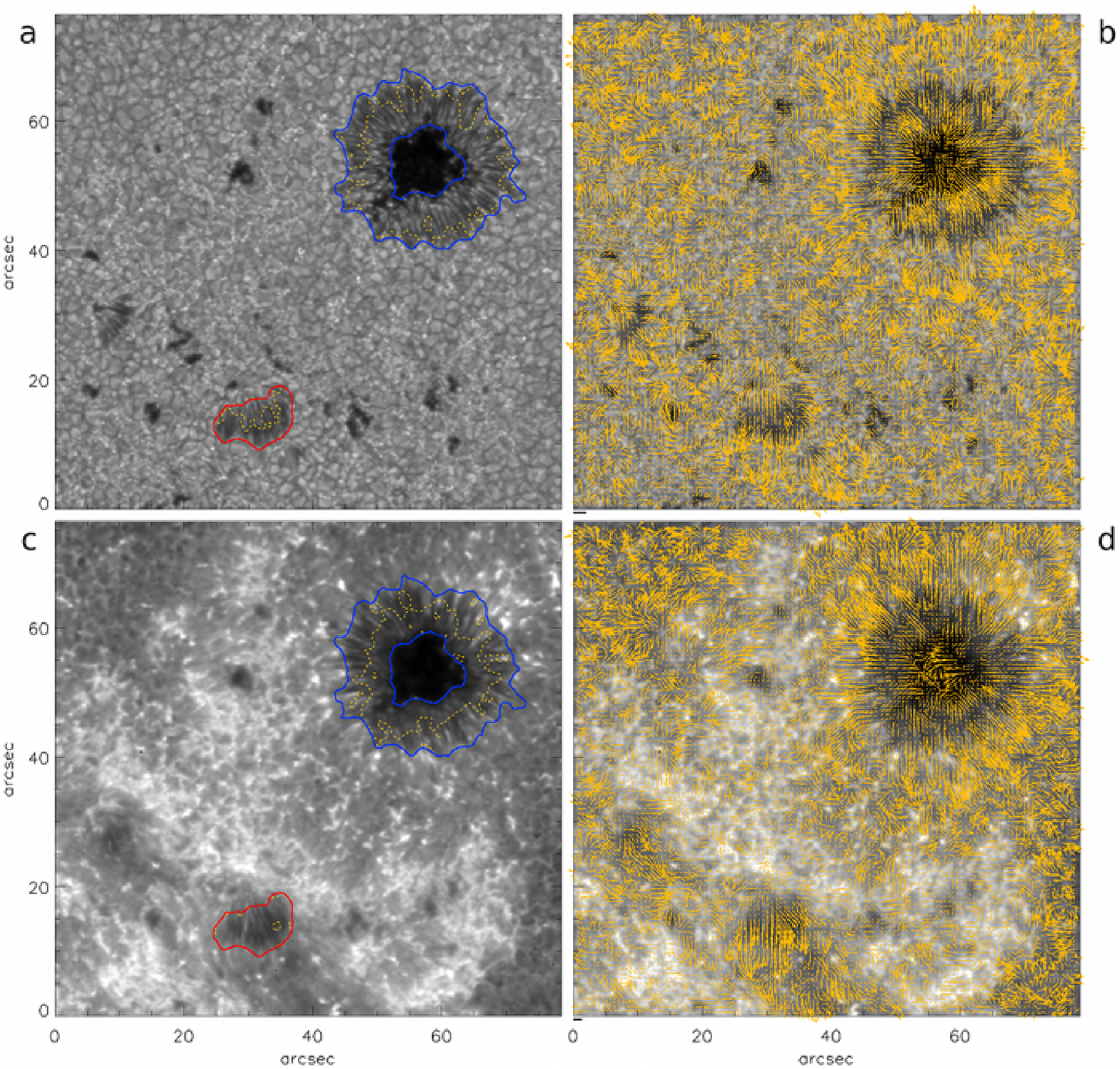}
 \caption{The G-band  and \ion{Ca}{II}~H filtergrams of active region NOAA~10960 taken at
 15:51~UT on 6 June 2007 (a, c). The red and blue
 contours mark the position of orphan and regular penumbrae,
 respectively, and are based on smoothed G-band intensity images. The
 arrows show the apparent horizontal motions of structures in the
 G-band (b) and \ion{Ca}{II}~H (d) filtergrams. The thick lines below
 the beginning of the $x$-axes correspond to 1~km~s$^{-1}$. The yellow
 dashed contours in (a) and (c) mark the boundary between the inward
 and outward apparent motions seen in (b) and (d), respectively.}
 \label{flowmap}
\end{figure*}

\subsection{Brightness and apparent motions}
\label{filtergrams}

Figure~\ref{flowmap} shows a late phase of the evolution of an
orphan penumbra in the active region NOAA~10960. A sequence of images
taken $\pm9$~minutes around the G-band filtergram as displayed in
Fig.~\ref{flowmap}a, was used to derive apparent horizontal motions,
$\vec{U}_{LCT}$, in the low photosphere (Fig.~\ref{flowmap}b).
Figure~\ref{flowmap}c shows the \ion{Ca}{II}~H filtergram co-spatial
and co-temporal with Fig.~\ref{flowmap}a. Arrows in
Fig.~\ref{flowmap}d represent the apparent motions in the upper
photosphere derived from the 18~minutes sequence of \ion{Ca}{II}~H
images. The arrows are shortened by a factor of two compared to
Fig.~\ref{flowmap}b. In Table~\ref{table_vel}, we give the mean values
of $|\vec{U}_{LCT}|$ and their standard deviations in different parts
of the G-band and \ion{Ca}{II}~H FOVs.

\begin{figure*}[!t]
 \centering \includegraphics[width=0.99\linewidth]{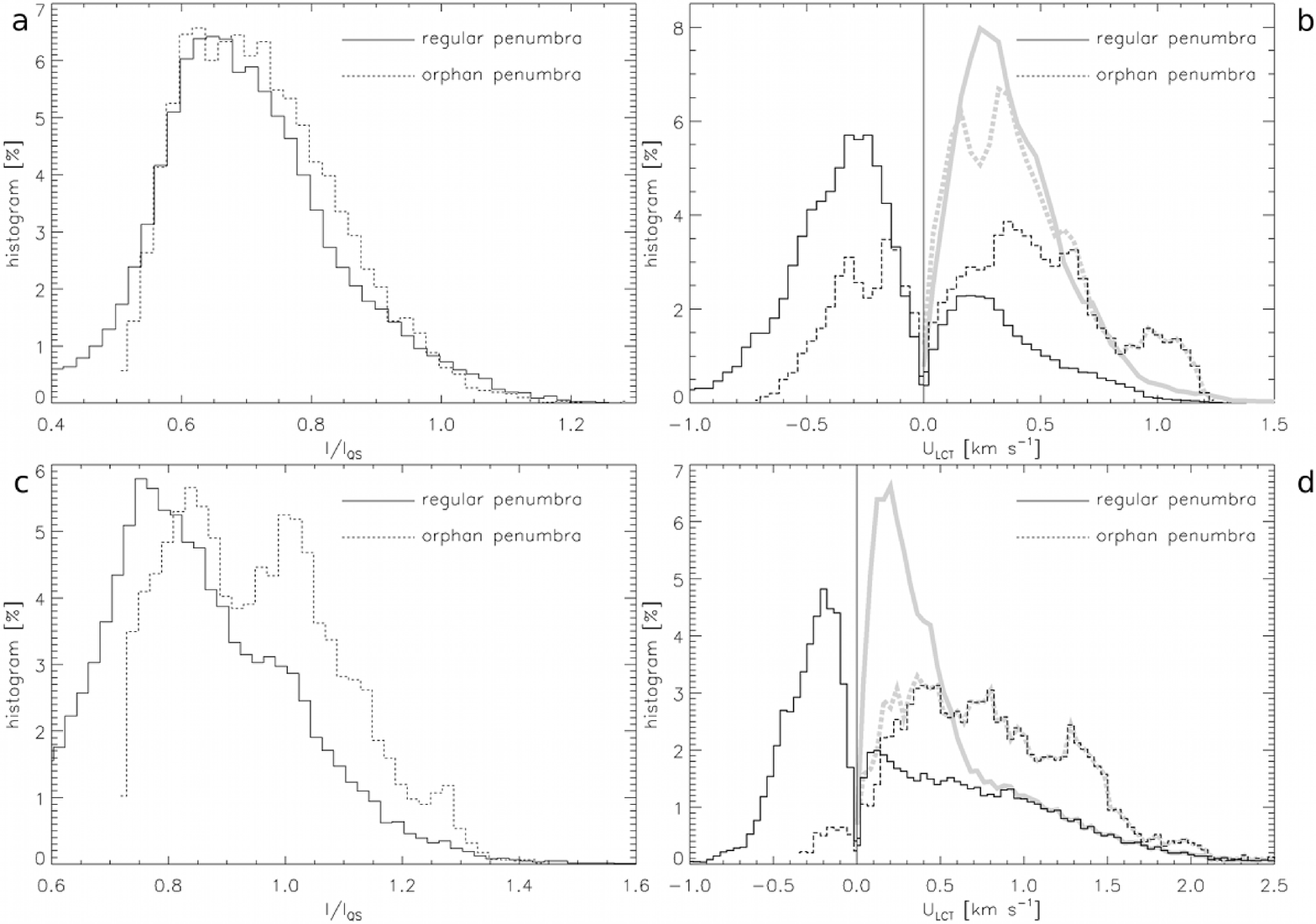}
 \caption{(a) and (b) Histograms of the G-band intensity and apparent
 horizontal velocities of pixels encircled by contours in
 Fig.~\ref{flowmap}a. (c) and (d) Histograms of the \ion{Ca}{II}~H
 intensity and apparent horizontal velocities of pixels encircled by
 contours in Fig.~\ref{flowmap}c. Inward motions have negative sign. The
thick grey lines in the histograms of apparent 
horizontal velocities show the distribution of $|\vec{U}_{LCT}|$.} \label{histograms}
\end{figure*}

\begin{table}[!t]
  \caption{Apparent horizontal velocities.} \centering
  \begin{tabular}{lcccc} \hline\hline \noalign{\smallskip} Region &
  \multicolumn{2}{c}{$\overline{|\vec{U}_{LCT}|}$ [km~s$^{-1}$]} & \multicolumn{2}{c}{$\sigma_{|\vec{U}_{LCT}|}$
  [km~s$^{-1}$]} \\ & G-band & \ion{Ca}{II}~H & G-band &
  \ion{Ca}{II}~H \\ \hline Regular penumbra & 0.41 & 0.59 & 0.25 &
  0.68 \\ \hfill inflow & 0.42 & 0.32 & 0.25 & 0.51 \\ \hfill outflow
  & 0.38 & 0.83 & 0.24 & 0.73 \\ \noalign{\smallskip} Orphan penumbra
  & 0.45 & 0.86 & 0.28 & 0.55 \\ \hfill inflow & 0.29 & 0.18 & 0.15 &
  0.10 \\ \hfill outflow & 0.53 & 0.89 & 0.29 & 0.55 \\
  \noalign{\smallskip} Moat region (RP) & 0.57 & 0.78 & 0.31 & 0.44 \\
  Moat region (OP) & 0.54 & 0.69 & 0.23 & 0.44 \\ \noalign{\smallskip}
  Quiet Sun & 0.57 & 0.86 & 0.31 & 0.46 \\ \hline \end{tabular}
  \tablefoot{Mean values $\left(\overline{|\vec{U}_{LCT}|}\right)$ and standard deviations
$\left(\sigma_{|\vec{U}_{LCT}|}\right)$ of apparent horizontal velocities   in various regions of the solar atmosphere as
determined from the
  G-band and \ion{Ca}{II}~H images. The abbreviations RP and OP stand for regular and orphan penumbra,
respectively.} \label{table_vel}
\end{table}

As can be seen in Fig.~\ref{flowmap}a, the morphology of orphan and
regular penumbrae is very similar in the lower photosphere. Both have
filaments with bright heads and darker tails. Figure~\ref{histograms}a
shows histograms of the G-band intensities in the two types of
penumbrae. The shapes of the histograms are similar. The dark pixels
with $I < 0.5 I_{QS}$ in the regular penumbra are a simple consequence
of the umbra/penumbra boundary, which does not exist in the orphan
penumbra. In both cases, the most frequent intensities are around
$0.65 I_{QS}$. The excess of pixels with intensities between $0.7
I_{QS}< I <0.9 I_{QS}$ in the orphan penumbra is partially caused by
the bright regions near the outer boundary of the filamentary
structure, which are proportionally larger compared to the regular
penumbra. Even if these boundary regions are not taken into account,
there is an excess of bright pixels in the orphan penumbra (not shown in
Figure~\ref{histograms}a). On average, the orphan penumbra is slightly brighter
than the regular penumbra when observed through the Hinode G-band filter.

In the \ion{Ca}{II}~H images, the orphan and regular penumbrae also
have a similar appearance (Fig.~\ref{flowmap}c). However, a quantitative
analysis reveals some differences. The histogram of \ion{Ca}{II}~H
intensities in the regular penumbra has only one peak at $0.75 I_{QS}$
(Fig.~\ref{histograms}c, solid line). For the orphan penumbra, the
main peak is located at around $0.85 I_{QS}$. There is a second peak
around $1.0 I_{QS}$, which is created by pixels near the orphan
penumbra boundary. A hint of bright pixels near the outer edge of the
regular penumbra can be seen as a bump of the solid line in
Fig.~\ref{histograms}c. Due to the different sizes of the two
penumbrae, the boundary regions show up more prominently in the orphan
penumbra histogram. Even if these regions are removed from the
histograms, the \ion{Ca}{II}~H intensities of the filaments are not
the same in the regular and orphan penumbrae (not shown in
Figure~\ref{histograms}c). The orphan penumbra is 
brighter by approximately $0.1 I_{QS}$, the intensity difference
between the main peaks of Fig.~\ref{histograms}c.

In Fig.~\ref{flowmap}b, we show the apparent flows found from the
G-band filtergrams. The motion pattern in the regular penumbra
confirms previous studies: In the inner penumbra, there are motions
towards the sunspot umbra, while one observes
outward motions in the outer penumbra  \citep[see][]{Wang:1992, Sobotka:1999,
  Marquez:2006}. Hereafter, motions opposite to the Evershed flow are referred to as inflows (inward motions) and flows in
the direction
of the Evershed flow as outflows (outward motions). The
$\vec{U}_{LCT}$ pattern in the orphan penumbra resembles that of the
regular penumbra. Everywhere in the orphan penumbra, the apparent
motions are aligned with the filaments. In regions corresponding to
the inner penumbra, there are inward apparent motions (upper region
of the orphan penumbra in Fig.~\ref{flowmap}).  In regions
corresponding to the outer penumbra, $\vec{U}_{LCT}$ points outwards.

In Fig.~\ref{histograms}b, we show the distribution of apparent
horizontal motions as derived from the sequence of G-band images.  The
regular penumbra histogram made from all pixels (solid grey line) resembles the one found
by \citet{Marquez:2006} in a sunspot penumbra. The histogram peaks
around 0.2~km~s$^{-1}$ with maximum values reaching 1.5~km~s$^{-1}$.
The mean value of $|\vec{U}_{LCT}|$ in the regular penumbra is
0.41~km~s$^{-1}$, which is about 0.1~km~s$^{-1}$ lower than the values obtained
by \citet{Marquez:2006} and \citet{Wang:1992}. The distribution
of apparent motions of all pixels in the orphan penumbra is shown by dashed grey line in
Fig.~\ref{histograms}b. Like in the regular penumbra, the most
frequent velocities are around 0.3~km~s$^{-1}$ with a mean value of
0.45~km~s$^{-1}$. The main difference compared to the regular penumbra
is the number of pixels with velocities around 1~km~s$^{-1}$.  These
flows are associated with outward motions (dashed black line - positive
values) having a mean apparent velocity of
0.53~km~s$^{-1}$, which is faster than in the regular penumbra
(0.38~km~s$^{-1}$, solid black lines - positive values). On the other
hand, the mean inward $\vec{U}_{LCT}$ is 0.29~km~s$^{-1}$ (dashed black
line - negative values), which is slower than in the regular penumbra
(0.42~km~s$^{-1}$, solid black lines - negative values, see also
Table~\ref{table_vel}). This might be partially explained by the position of the
contour around the orphan penumbra, which does not encircle most of the bright
penumbral grains whose motions give rise to inward flows.

The apparent motions derived from the \ion{Ca}{II}~H filtergrams are
presented in Fig.~\ref{flowmap}d. The pattern in the regular penumbra
is the same as in Fig.~\ref{flowmap}b, which shows inward motions in the
inner penumbra and outward motions in the outer penumbra. However,
there is an important difference between the areas occupied by inward
and outward motions in the G-band and \ion{Ca}{II}~H maps. The yellow
dashed contours in Figs.~\ref{flowmap}a and~\ref{flowmap}c mark the
boundary between oppositely oriented flows.  This boundary moves
significantly towards the sunspot umbra in the \ion{Ca}{II}~H map. The
inward flows are determined by the LCT method from the apparent
motions of bright penumbral grains.  As these structures are confined
to the low photosphere \citep{Jurcak:2007, Sobotka:2009}, they are
prominent in the G-band filtergrams but not in the \ion{Ca}{II}~H
images.  Therefore, the inward motions are weaker and occupy smaller
areas in \ion{Ca}{II}~H. The motion patterns of the orphan
penumbra are also similar in the upper and lower photosphere
(Figs.~\ref{flowmap}d and \ref{flowmap}b). The inward motions almost
completely disappear in the orphan penumbra for the same reason as in
the regular penumbra. This also occurs because the orphan penumbra
boundary (red contour in Figs.~\ref{flowmap}a and~\ref{flowmap}c) does
not include the area where the bright penumbral grains are most
common.

The distributions of apparent motions derived from the sequence of 
\ion{Ca}{II}~H images are, however, significantly different in regular 
and orphan penumbrae, as can be seen in Fig.~\ref{histograms}d. In the
regular penumbra (solid grey line), there is a main peak at 0.2~km~s$^{-1}$, which is
mostly created by inward motions (with mean $\vec{U}_{LCT}$ of
0.32~km~s$^{-1}$, solid black line - negative values).  The outward
flows have a flat distribution with 
mean velocities of 0.83~km~s$^{-1}$ and maximum values of
2~km~s$^{-1}$ (solid black line - positive values). In the orphan
penumbra, there are almost no regions 
with inflows (The mean value of the inflow $\vec{U}_{LCT}$ is 0.18~km~s$^{-1}$
with a maximum values of 0.4~km~s$^{-1}$, which is shown as a dashed black line - negative
values.). The distribution of outflows 
is similar in the orphan and the regular penumbra, which is a flat
distribution with maximum velocities over 2~km~s$^{-1}$ and mean flows
of 0.89~km~s$^{-1}$ (dashed black line - positive values). Taking into
account all the pixels in the regular and 
orphan penumbrae, the mean apparent motions are 0.59~km~s$^{-1}$ and
0.86~km~s$^{-1}$, respectively (see Table~\ref{table_vel}). The flows
are faster compared to the lower atmospheric layers due to the
significant increase of apparent velocity of the outward flows.

In both the G-band and \ion{Ca}{II}~H horizontal flow maps, one can
distinguish a moat-flow region around the main sunspot except for the
lower left area, where the penumbra is not fully developed. This agrees with recent studies of the moat flow
\citep{Vargas:2007,Vargas:2008}. The apparent motions in the
orphan penumbra also extend their boundaries in the direction of the flow,
creating a structure comparable to a sunspot moat region. In the deep
photosphere, the apparent motions exhibited by the orphan penumbra
``moat'' region have amplitudes comparable to the outward motions
found inside the penumbra, while the moat flows around the sunspot
have higher velocities than their penumbral counterparts (see
Table~\ref{table_vel}). In the upper photosphere, the velocities of
the apparent motions detected in the moat regions are smaller than
those found in the penumbra, although the difference is small in the
case of the regular penumbra.  The absolute values of the moat
$\vec{U}_{LCT}$ increase as one moves higher in the atmosphere
\citep[see Table~\ref{table_vel} and also][]{Sobotka:2007}.

\begin{figure*}[!t]
 \centering \includegraphics[width=0.95\linewidth]{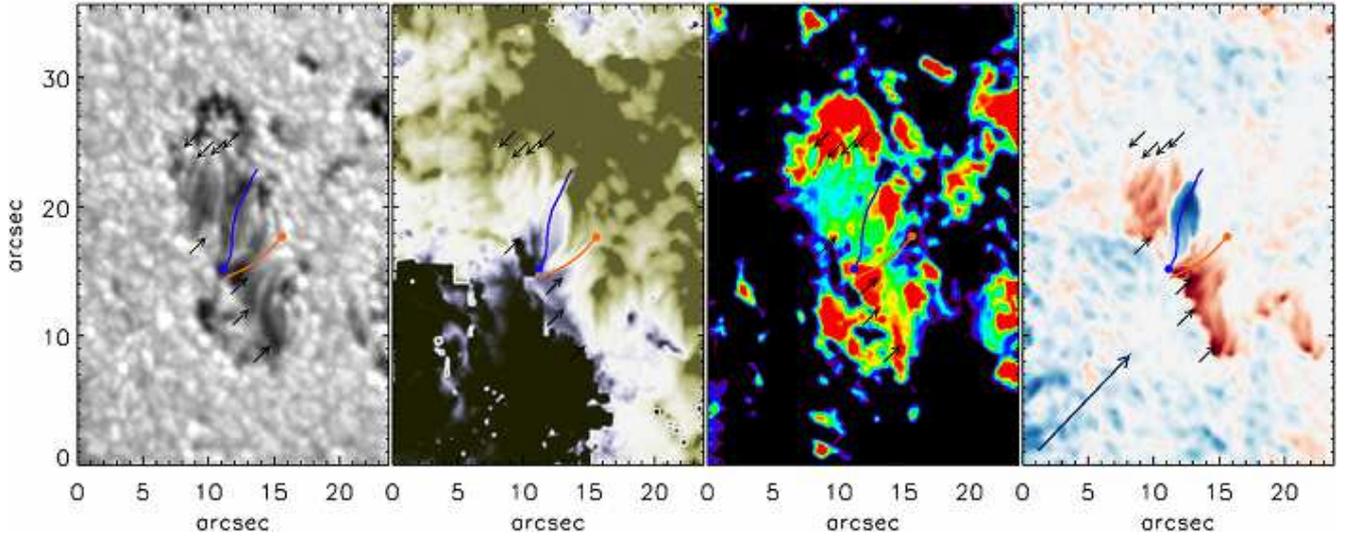}
 \caption{Examples of orphan penumbral filaments in area no.~2. The
 panels are identical here and in Fig.~\ref{detail_op}. The small arrows
 pointing up and down (in the lower and upper parts of the maps) mark
 the tails and heads of selected filaments, respectively. The blue and
 orange lines follow filaments with oppositely oriented flows. The large
 arrow points toward the disc centre.}	
 \label{siphon_map}
\end{figure*}

\section{Discussion}
\label{discussion}

As shown in Sect.~\ref{results}, there are many similarities between
regular and orphan penumbrae. Their filaments look identical in the
atmospheric layers probed by the G-band and \ion{Ca}{II}~H images.  In
the deep photosphere, orphan penumbrae are only slightly brighter than
regular penumbrae, while the intensity difference is significant in
the upper photosphere, as demonstrated by the \ion{Ca}{II}~H
filtergrams. Such an intensity difference could be explained
by the filaments reaching different heights in orphan and regular
penumbrae. The background component of the magnetic field in the
latter creates a magnetic canopy \citep{Borrero:2008a} that may limit
the geometrical height attainable by the flow channels containing the
hot gas, whereas the former would not be impeded in any way.

\begin{figure*}[!t]
 \centering 
 \includegraphics[width=0.99\linewidth]{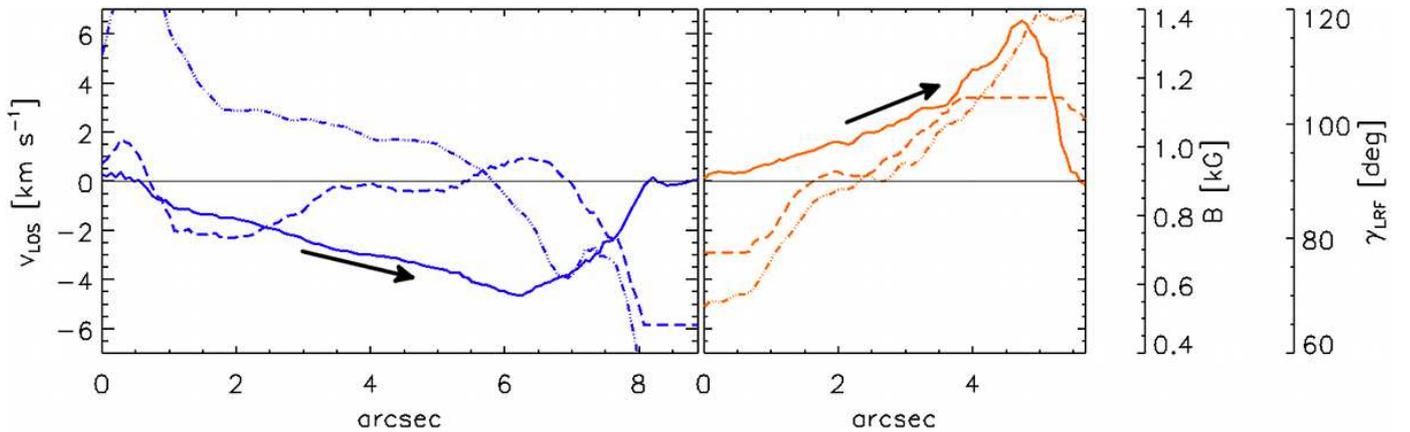}
 \caption{Variation of magnetic field strength (dashed lines), LRF inclination (dash-dotted lines), and
 LOS~velocity (solid lines) along the filaments marked by orange and blue
 lines in Fig.~\ref{siphon_map}. Distances are measured from the
 filament heads (circles in Fig.~\ref{siphon_map}). The black arrows
show the direction of the flows along the cuts.}
 \label{siphon_plot}
\end{figure*}

Despite the possible height difference, the filaments of orphan and
regular penumbrae share a common feature: their heads (bright
penumbral grains) and tails are low-lying structures visible only in
the G-band filtergrams. When we compare the \ion{Ca}{II}~H with the 
G-band images, the filaments appear shorter in the upper photosphere
and do not reach the boundaries based on G-band intensities
(Figs.~\ref{flowmap}a and~\ref{flowmap}c). This agrees with
the magnetic configuration we have deduced for the orphan penumbrae,
where the magnetic field is surfacing and submerging at the heads and
tails of the filaments. The shortest filaments on the edges of orphan
penumbrae cannot be seen in the
\ion{Ca}{II}~H filtergrams. There is also a small segment of orphan
penumbra located around [10, 30]~arcsec in Figs.~\ref{flowmap}a
and~\ref{flowmap}c, which is obviously smaller in the \ion{Ca}{II}~H
image compared to the G-band filtergram.

The general properties of the observed apparent horizontal motions are
similar in orphan and regular penumbrae, which are inward motions near the
filament heads and outward motions around the tails. The shapes
of the histograms of inflows and outflows are also comparable. As shown by
\citet{Vargas:2008}, the moat flow seems to be closely related to the
presence of the Evershed flow. Therefore, it is interesting to compare
the mean outflow velocities found in orphan and regular penumbrae with
the flows observed in their moat regions (see Table~\ref{table_vel}).
The moat flows are faster around the regular penumbra in the two
photospheric layers probed by the observations, although the outward
motions are faster in orphan penumbrae and we also find faster
``Evershed'' flows there. Such a discrepancy may support the
  results of \citet{Lohner:2013}, who found no correlation between the
  Evershed and moat flow velocities and interpreted this as a hint
  that the two flows have different physical origins. However, the
  presence of moat flows on the sides of orphan penumbrae towards 
  which the apparent motions are directed supports the conclusions of
  \citet{Vargas:2008}.

We cannot directly confirm the horizontal direction of the gas flow in
orphan penumbrae.  However, the LOS~velocity measured in the orphan
and regular penumbrae of active regions NOAA~10960 and 11089 show the
same decreasing trend as they approach the central meridian. Some orphan 
penumbral filaments are also almost perpendicular to the
direction towards disc centre and exhibit very small LOS~velocities.
Since the magnetic field in the filaments
is mostly horizontal, it is reasonable to assume that the flow is
significantly inclined. Observations of NOAA~10960 taken close to the
disc centre show fast downflows at the end of orphan penumbral
filaments, which are analogous to the downflows observed in the outer
parts of regular penumbrae. There is no clear evidence of upflows in
bright grains of the orphan penumbra, but these are also not
detectable in the nearby regular penumbra (possibly due to 
insufficient spatial resolution).

To understand the origin of the gas flows, we need to examine how they
change along penumbral filaments. The arrows pointing down in
Fig.~\ref{siphon_map} mark the heads of a few selected filaments, where
the flow appears (rightmost panel). These regions are co-spatial with
spines that have weaker and more horizontal magnetic field (middle panels)
and with bright penumbral grains (leftmost panel). There is a slight
offset between the arrow heads and the bright grains in the continuum
intensity map. As mentioned in Sect.~\ref{observations}, the maps of
physical parameters correspond to a geometrical height around 150~km,
while the continuum intensity is formed around 0~km. This height
difference with the location approximately 40$^\circ$ from
the disc centre can explain the observed offset. The arrows pointing
up mark the regions, where the filaments submerge below the solar
surface. These areas are associated with stronger flows (left map) and
correspond to regions with locally stronger magnetic fields pointing
towards the solar surface (middle maps).

The blue and orange lines in Fig.~\ref{siphon_map} mark two individual
filaments with oppositely oriented flows. They have been tracked
manually using the continuum intensity, LOS~velocity, and magnetic
field inclination of the structures. Figure~\ref{siphon_plot} shows
the LOS~velocity ($v_\mathrm{LOS}$), magnetic field strength ($B$),
and LRF inclination ($\gamma_\mathrm{LRF}$) along these paths.  The
blue lines correspond to the cut through the filament with
blueshifts. The flow forms in regions with $B$ around 750~G and
$\gamma_\mathrm{LRF}$ around 110$^\circ$ (negative polarity
patch). There is a gradual increase of the LOS~velocity up to
$-5$~km~s$^{-1}$, where the magnetic field strength reaches 950~G and
turns towards the solar surface with an inclination of around
80$^\circ$ (positive polarity patch).  In the filament with
redshifts, we also find a flow towards a stronger magnetic field. The flow
forms in regions with $B$ of 650~G, while the maximum value of 1400~G
is reached at the location with the highest LOS~velocity of
6.5~km~s$^{-1}$.  The filament shown in red starts in a
positive polarity patch ($\gamma_\mathrm{LRF} = 70^\circ$) and
submerges in a negative patch ($\gamma_\mathrm{LRF} =
120^\circ$). This configuration of the flow and the magnetic field
clearly indicates a siphon flow.

The siphon flow mechanism was suggested as a possible driver of the
Evershed flow by \citet{Meyer:1968}: the difference of magnetic field
strength at the footpoints of a loop results in different gas
pressures, which generates a mass flow towards the footpoint with stronger
magnetic field \citep[][]{Thomas:1988, Degenhardt:1991}. This seemed
to be an unrealistic scenario for sunspot penumbrae because most analyses
showed significantly weaker magnetic fields in the outer
penumbra. Some studies suggest that the field can be stronger than the
background field component \citep{Bellot:2004,Tritschler:2007,
Ichimoto:2008, Borrero:2008}, and there are analyses that support the
siphon flow mechanism indirectly \citep{Borrero:2005,
Montesinos:1997}. More recent studies using complex inversion
schemes found strong magnetic fields co-spatial with submerging
Evershed flow channels in sunspot penumbrae
\citep{Tiwari:2013, Noort:2013}. \citet{Rempel:2011} made the 
suggestion that the driver of the flow could be the strong horizontal
component of the Lorentz force. He also stated that siphon-like flow
channels can be produced by such a process in the outer penumbra but
that they are more a consequence of the fast outflow than its
cause. Our analysis does not allow us to decide if the magnetic
configuration we have found in orphan penumbrae is the source or the
consequence of the fast flow.

Taking into account all the similarities between orphan and
  regular penumbrae, we propose that orphan penumbral filaments
  represent the flow channels of regular penumbrae, which are not surrounded or
  confined by a stronger and more vertical background field. This is
  also supported by the studies of \citet{Lim:2013} and
  \citet{Zuccarello:2013}, who found no changes in the structure of
  chromospheric magnetic fields during the formation of orphan
  penumbrae.

The magnetic configuration we have deduced for orphan penumbrae is
that of a flat $\Omega$-loop. The temporal evolution of the
opposite-polarity regions surrounding them suggests that the
disappearance of orphan penumbrae is caused by the submergence of the
$\Omega$-loop.  However, this process is very slow, and we do not
detect the corresponding Doppler shifts. Even the observations of
NOAA~10960 at an heliocentric angle of 8$^\circ$ did not reveal
downflows in the horizontal part of the filaments.  Assuming that the
height of the structure creating the orphan penumbra is comparable to
its width, a disappearance time of 30 hours would result in a
submergence velocity of 40~m~s$^{-1}$. Due to the horizontal gas flows,
the local variations of the LOS velocity observed within the orphan
penumbrae are much larger than this, making it impossible to detect
the small downward motion of the loop as a whole.

Our interpretation that the disappearance of orphan penumbrae is
the consequence of the submergence of an $\Omega$-shaped magnetic loop
does not contradict with \citet{Lim:2013} and
\citet{Zuccarello:2013}. They find that orphan penumbrae are created
by emerging $\Omega$-loops that are trapped in the photosphere by
an overlaying chromospheric magnetic field but did not discuss 
the cause of orphan penumbral disappearance in detail.

\section{Conclusions}
\label{conclusions}

We thoroughly investigated orphan penumbrae, which are filamentary structures
resembling sunspot penumbrae but are not connected to an umbra that lie
near PILs in active regions.  We observed these structures in two
active regions and used Hinode SOT data to analyse their properties.

In the G-band, orphan and regular penumbrae have the same morphology,
although the former are slightly brighter on average. The intensity
difference is more pronounced in the upper photosphere, which is probed by the
\ion{Ca}{ii}~H filtergrams (Figs.~\ref{histograms}a
and~\ref{histograms}c).  We suspect the reason for the enhanced
intensity of the filaments in orphan penumbrae is their more elevated
geometrical height, as they are not limited by the magnetic canopy of
the background field present in sunspots.

Although there are some differences in the amplitudes of apparent
motions in orphan and regular penumbrae, their overall structure is
the same in the low and upper photosphere.  We observe apparent motions
opposite to the gas flow in the inner penumbra (heads of the
filaments) and apparent motions following the flow in the outer parts
of orphan and regular penumbrae (Figs.~\ref{histograms}b
and~\ref{histograms}d). Similarly to sunspots, there is a moat region
near orphan penumbrae, which is located in the area towards which the flow is
directed (extending the relation between the Eveshed flow and moat region found by
\citealt{Vargas:2007} to orphan penumbrae).

The analysis of Hinode SP and NFI data shows that orphan penumbrae are
magnetic structures with the shape of flat $\Omega$~loops. During the evolution of orphan penumbrae,
the opposite polarities flanking them approach each other, while the
filaments get shorter and disappear, which points to the submergence
of the whole structure. Our velocity measurements cannot be used to
demonstrate the downward motion as they are dominated by ``Evershed''
flows, and even observations taken close to the disc centre do not
exhibit downward motions. We also cannot distinguish whether the
filaments disappear earlier in the \ion{Ca}{II}~H images or the
G-band filtergrams.

The LOS velocities observed in orphan penumbrae are larger than 
those of nearby regular penumbrae, but the structure of the flow is 
the same. The lowest velocities occur near the filament heads, while the flow
speed increases along the filaments and reaches a maximum in the tails.
Using full Stokes polarimetry, we were able to determine the
orientation and strength of the magnetic field along individual
filaments of orphan penumbrae (Figs.~\ref{detail_op}, \ref{sp_evol},
\ref{siphon_map}, and~\ref{siphon_plot}). We found that the flows are
oriented towards the footpoints with a stronger magnetic field, which agrees 
with the siphon flow mechanism as the driver of the flow.

Due to the similarities between orphan and regular penumbrae, we
tentatively propose that orphan penumbrae represent the weaker and
more inclined magnetic field component of regular penumbrae \citep[not
influenced by the background field present in sunspots,][]{Solanki:1993}.  On the other hand, this paper supports the
siphon flow mechanism as the driver of the Evershed flow.  It remains
to be studied if the observed magnetic field configuration is just a
consequence of the fast flow \citep[as suggested by][]{Rempel:2011} or
the actual driver of the flow.

\begin{acknowledgements}

The support from GA~CR~P209/12/0287 and RVO:67985815 is gratefully
acknowledged.  This work has been funded by the Spanish MINECO through
project AYA2012-39636-C06-05, including a percentage from European
FEDER funds.  Hinode is a Japanese mission developed and launched by
ISAS/JAXA, with NAOJ as domestic partner and NASA and STFC (UK) as
international partners. It is operated by these agencies in
cooperation with ESA and NSC (Norway).

\end{acknowledgements}

\bibliographystyle{aa}
\bibliography{muj}

\end{document}